\documentclass{is2014}

\title{Ultra-peripheral Collisions at RHIC: An Experimental Overview}

\author{Spencer R. Klein \\
Lawrence Berkeley National Laboratory, Berkeley, CA, USA}

\date{\today}

\begin{document}
\maketitle
\thispagestyle{fancy}

\begin{abstract}
Ultra-peripheral collisions (UPCs) of ions allow us to study photonuclear and two-photon interactions at energies above those available at fixed target accelerators.  For heavy ions, the couplings are large enough so that multi-photon interactions are possible, and higher order corrections are expected to be significant.  In this writeup, I present some recent UPC results from the Relativistic Heavy Ion Collider (RHIC), and discuss some future prospects.  I also draw parallels between UPC data and that expected at an electron-ion collider (EIC), and show how UPCs are a useful lead-in to EIC physics.  This writeup is based on a talk at "Initial State 2014," (IS2014), with a focus on the newest results.  One important result is that comparison of the RHIC (and LHC) results on coherent $\rho^0$ photoproduction show evidence for nuclear suppression, compared to a calculating based on $\gamma p$ cross-sections. 

\end{abstract}

\maketitle


\section{\label{sec:intro}Introduction}

Ultra-peripheral collisions occur when two ions pass by each other with an impact parameter ($b$) large enough so that they do not interact hadronically, but small enough so that electromagnetic interactions occur \cite{Baur:2001jj,Bertulani:2005ru}.  The photon flux comes from the electromagnetic fields of the colliding ions.  Both $\gamma\gamma$ and photonuclear interactions are possible.   $\gamma\gamma$ interactions probe quantum electrodynamics in strong fields, while photonuclear interactions are sensitive to the structure of the target nucleus.  The latter topic is a key focus of the Initial State
2014 conference. 

There are several reasons to study UPCs at heavy ion colliders.  The photon flux scales as $Z^2$, so interaction rates can be very high, and it is possible to exchange more than one photon in a single ion-ion encounter.   The maximum photon energy is $\Gamma\hbar c/R_A$, where $\Gamma$ is the Lorentz boost of the ion (in the target frame).
At RHIC, the maximum target-frame photon energy is about 600 GeV.  

Multi-photon interactions are an important experimental tool for studying UPCs.  Photoproduction (or $\gamma\gamma$ interactions) accompanied by mutual Coulomb excitation is of particular interest.  One photon produces a vector meson or other final state, while the two nuclei exchange two additional photons, each of which excites the other nucleus.  When the two excited nuclei decay, they emit neutrons which provide a convenient experimental trigger \cite{Adler:2002sc}.  Each photon is emitted independently, linked only by a common impact parameter  \cite{Baur:2003ar}, and the cross-sections can be computed by integrating over the impact parameter \cite{Baltz:2002pp,Baltz:2009jk}
\begin{equation}
\sigma = \int d^2\vec{b} P_1(b) P_2(b)P_3(b)...P_{\rm Nohad}(b)
\end{equation}
where $P_1(b) $ is the probability for reaction 1 to occur at impact parameter $b$ and $P_{\rm Nohad}(b)$ is the probability not to have a hadronic interaction, which would obscure the UPC.  Away from the kinematic limits, $P(b)_i\propto 1/b^2$.   For Coulomb excitation, these probabilities may be large, and so may require unitarization: $P_1(b) = 1 -\exp[-P_1'(b)]$, where $P_1'(b)$ is the non-unitarized probability.   Besides allowing a simple trigger, multi-photon interactions occur at considerably smaller average impact parameters than single-photon exchange, leading to a considerably harder photon spectrum.  Also, as we will see below, the common $\vec{b}$ introduces correlations between the final states. 

\section{Two-photon interactions}

Although the cross-sections for most $\gamma\gamma$  interactions are lower than for photonuclear interactions \cite{Bertulani:2005ru}, lepton pairs are copiously produced.  The total cross-section for $e^+e^-$ pairs with gold beams at RHIC is 100,000 barns!  Most of these leptons have very low transverse momentum, $p_T \approx m_e$, so they are not observed in a central detector.   However, enough of them have enough  $p_T$ so that they can be a significant background in vertex detectors. 
Both STAR and PHENIX have studied production of $e^+e^-$ pairs, although they only detect leptons with considerably higher $p_T$.  STAR studied pairs with lepton $p_T>55$ MeV/c accompanied by mutual Coulomb excitation \cite{Adams:2004rz}.  The STAR analysis found that the cross-section was compatible with the predictions of lowest order quantum electrodynamics (QED).  A later study using a newer theoretical calculation found that the result was more compatible with an all-order calculation, incorporating Coulomb corrections \cite{Baltz:2007gs}.  PHENIX studied high-mass $e^+e^-$ production as part of their analysis of $J/\psi$ photoproduction, and found results that were consistent with QED \cite{Afanasiev:2009hy}.

Another important $\gamma\gamma$ interaction is bound-free pair production, which happens when an $e^+e^-$ pair is produced with the electron bound to one of the ions.  The resulting single-electron ions have a reduced charge to mass ratio, but an unchanged momentum, so their trajectories gradually diverge  from the beam of bare ions \cite{Klein:2000ba}.  For gold, the beam of single-electron ions spreads out before hitting the beampipe.  However, for copper, the single-electron beam remains focused, striking the beampipe about 136 m downstream from the interaction region.  The resulting showers have been observed using a series of PIN diodes (part of the RHIC radiation monitoring system), at about the expected cross-section \cite{Bruce:2007mx}.  This process is important at higher energy colliders, where the single-electron ion beam may carry enough energy to quench superconducting magnets.

\section{Photonuclear Interactions - vector meson production}

The most commonly studied photonuclear interaction is vector meson photoproduction.  One nucleus emits a photon, which fluctuates into a quark-antiquark pair ($q\overline q$), which then scatters elastically (but hadronically) emerging as a real vector meson; the cross-section depends on the size of the $q\overline q$ dipole.   This occurs via meson (at low energies) and Pomeron (in modern terms, a gluon ladder) exchange.  In perturbative QCD, the cross-section depends on the gluon density $g$ of the target nucleus, $\sigma\propto g^2(x,Q^2\approx m_q^2)$, where $m_q$ is the quark mass \cite{Adeluyi:2012ph}.  The Bjorken-$x$ of the gluon is determined by the kinematics, $\gamma x M_p = M_v/2 \exp(\pm y)$, where $M_p$ is the proton mass, $M_v$ is the vector meson mass, and $y$ is its rapidity.  Here, $\gamma$ is the Lorentz boost of one nucleus, in the lab frame.  The $\pm$ is because of the ambiguity as to which nucleus emitted the photon and which is the target.

The cross-section for coherent $\gamma A\rightarrow VA$ interactions may be calculated using a Glauber calculation \cite{Klein:1999qj,Frankfurt:2002sv}, using the results of a $\gamma p\rightarrow Vp$ calculation, or using data as input.   The Glauber calculation accounts for the possibility for a $q\overline q$ dipole to interact multiple times within the nucleus.  For small dipoles, the coherent cross-section (with $p_T < \hbar/R_A \approx 100$ MeV/c) is enhanced as $A^2$, while for larger dipoles, multiple interactions reduce the cross-sections.  For the $\rho^0$, the cross-section scales very roughly as $A^{5/3}$.  HERA has accurate data on vector meson photoproduction in the energy range relevant to RHIC \cite{HERAdata}.  This data can be used to check calculations of $\gamma p\rightarrow Vp$ or as direct input to a Glauber calculation.
By comparing RHIC data with ion targets to the HERA data, one can measure nuclear effects, such as shadowing, which reduce the cross-section compared with a model where each nucleon is independent \cite{Lappi:2013am}.
Saturation models also predict significant nuclear dependence \cite{Goncalves:2009kda}.

STAR has studied vector meson photoproduction in gold-gold collisions at energies of 62 \cite{Agakishiev:2011me}, 130 \cite{Adler:2002sc} and  200 GeV \cite{Abelev:2007nb}; the 200 GeV data is the most precise.  The mid-rapidity $d\sigma/dy$ is about half that predicted by a quantum Glauber calculation \cite{Frankfurt:2002sv}, but is consistent with the predictions of a classical Glauber calculation \cite{Klein:1999qj}.  Recent ALICE data on $d\sigma/dy$ for $\rho$ photoproduction in lead-lead collisions exhibits a similar factor of two suppression below the quantum Glauber calculation, and is also consistent with a classical Glauber approach \cite{ALICE}.  One likely explanation is that the $\rho$ production is reduced because of gluon shadowing or other nuclear effects.  Normally, one does not expect perturbative QCD phenomena such as shadowing to apply for vector mesons as light as the $\rho$, but that possibility needs to be considered.  

\begin{figure} [tb]
\center{\includegraphics[width=0.45\textwidth]{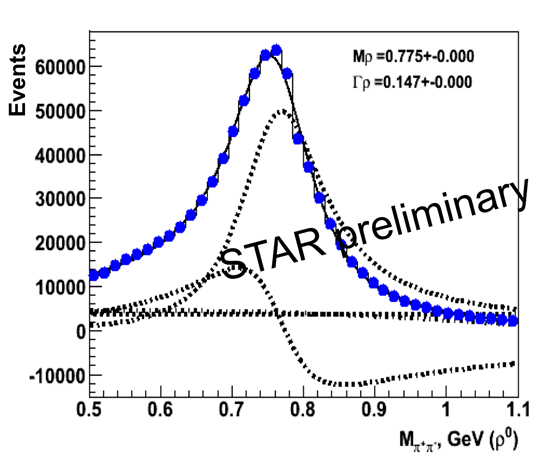}}
\caption{\label{fig:rhomass} The $\pi^+\pi^-$  invariant mass distribution for STAR UPC events.  The mass distribution is fit to a combination of $\rho$ and direct $\pi^+\pi^-$ photoproduction (dashed line with peak), plus an interference term (dashed oscillating line) and a background (flat line), determined from the like-sign events   From Ref. 
\cite{DNP}.
} 
\end{figure}

The latest STAR study found about 650,000 $\rho^0$ events in 37 million triggers collected during the 2010 run \cite{Debbe:2013mia,Janet}.  The $\pi^+\pi^-$ invariant mass plot, shown in Fig. \ref{fig:rhomass}, was fit to a Breit-Wigner resonance (for the $\rho$), a constant contribution (for direct $\pi^+\pi^-$ production), an interference term, and a background term.  The fit returns a mass and width consistent with the known values for the $\rho$, and the ratio of $\rho^0$ to direct pions is similar to that observed in $\gamma p$ collisions at HERA. 

STAR has also studied coherent photo-production of excited $\rho$ states \cite{Abelev:2009aa}, decaying to four charged pions. This should occur mostly through the $\rho(1450)$ and/or the $\rho(1700)$.  STAR observed a coherent enhancement at low $p_T$, consistent with a mixture of the two resonances.   The cross-section seems  consistent with predictions based on generalized vector meson dominance \cite{Frankfurt:2002wc}.

PHENIX has studied $J/\psi$ photoproduction \cite{Afanasiev:2009hy}.   They find a cross-section consistent with the quantum Glauber model.  The $p_T$ spectrum exhibits a mixture of coherent and incoherent production.  They have also observed
 photoproduction of $J/\psi\rightarrow\mu^+\mu^-$ in the forward region ($1.2 < y < 2.2$), with  higher statistics, but, as of yet, no cross-section determination \cite{PHENIX}.  STAR has also studied $J/\psi$ photoproduction, and sees a signal of about 100 events in the 2010 data \cite{DNP}.   They have also not released a cross-section, but the ratio of $J/\psi$ to $\rho$ production is slightly higher than theoretical expectations \cite{DNP}. 

\begin{figure} [tb]
\center{\includegraphics[width=0.6\textwidth]{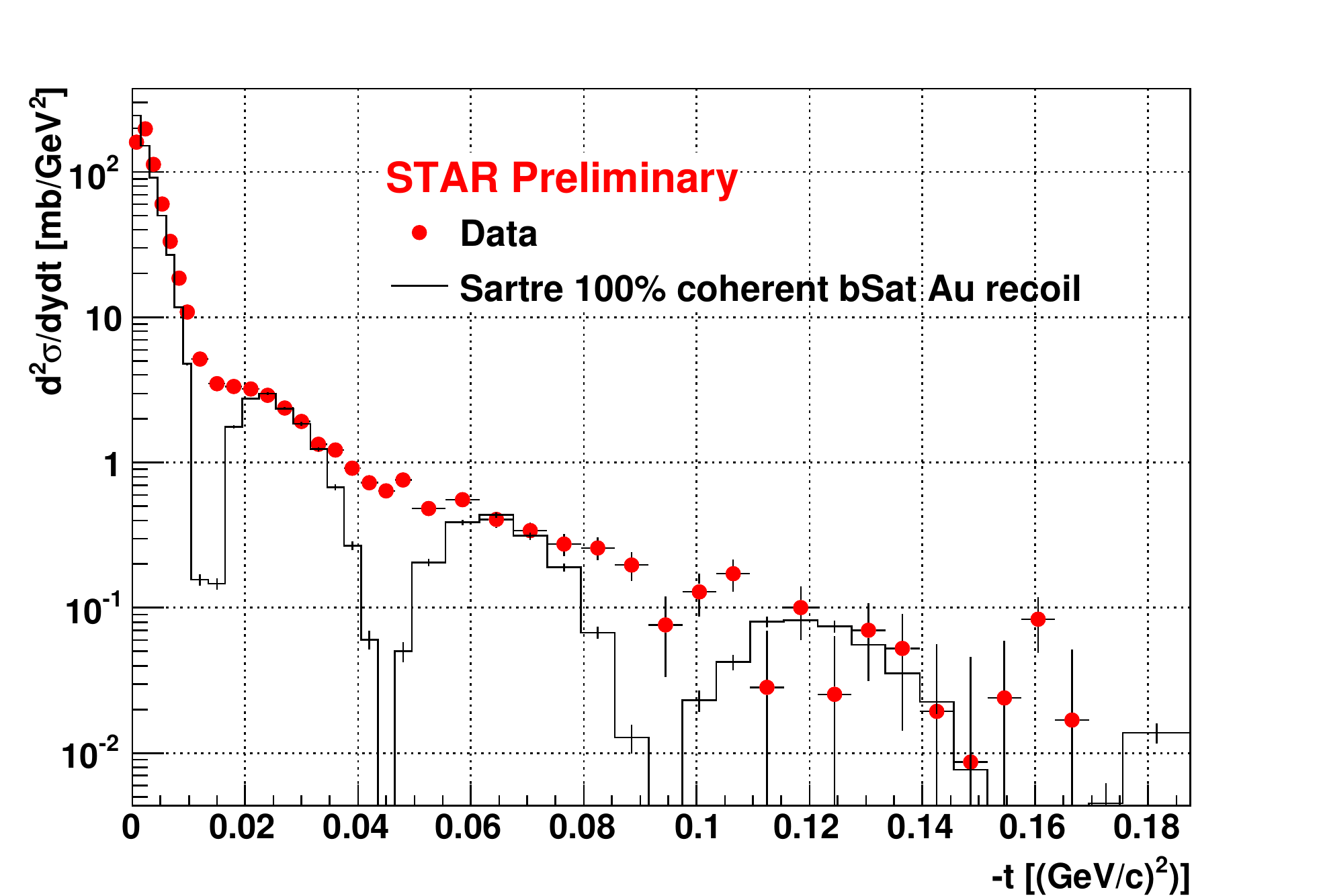}}
\caption{\label{fig:Ramirorho} $t_\perp = p_T^2$ distribution for photoproduced $\rho$ seen in STAR.  Two or three diffraction dips are visible.  The distribution is compared with predictions from the SARTRE Monte Carlo \cite{Toll:2012mb}, which reproduces the dip positions.   From Ref. \cite{Debbe:2013mia}.
\label{fig:rhopt}
} 
\end{figure}

Vector mesons may be produced via coherent or incoherent photoproduction. In coherent production, the nucleus remains intact, while with incoherent production, it breaks up.   However, in UPCs, the situation is complicated by the presence of additional photons, which can also cause the nucleus to break up.   Neutrons emitted by the nuclear breakup may provide a convenient signature to trigger a detector.  The STAR collaboration finessed this problem by selecting events with a single neutron in each zero degree calorimeter, under the expectation that most incoherent production generates more than one neutron.  The resulting $\rho$ $t_\perp = p_T^2$ spectrum is shown in Fig. \ref{fig:rhopt}, along with a comparison with the SARTRE Monte Carlo \cite{Toll:2012mb}.  Here, the longitudinal component of $t$ is small, so $t\approx t_\perp$. The spectrum is characteristic of diffraction, with at least two peaks visible, separated by a minima.  SARTRE predicts the position of the peaks correctly, but this version overestimates their depth, likely because it did not account for the $p_T$ of the emitted photon.  It is possible to Fourier transform the $p_T$ spectrum to determine the nuclear density profile, modulated by the effect of the Glauber superposition.  

\section{UPC geometry, polarization and final state particle correlations}

The geometry of UPC photonuclear collisions introduces interesting correlations between the particles produced in multi-photon interactions.   The electric field vectors point away from the emitting nucleus.  At the target nucleus, the electric field vectors of all of the photons are aligned with $\vec{b}$, so the photons share a common linear polarization.   If the two photons are both emitted from the same nucleus, then the electric fields point in the same direction, while if they are emitted by different nuclei, then the fields point in opposite directions.   

The linear alignment has experimental consequences, even for single particle production.  At mid-rapidity, each nucleus is equally likely to be a photon emitter or absorber. The electric fields for the two possibilities are anti-aligned, so the two production amplitudes can cancel out.  At $p_T = 0$, production is suppressed.  At finite $p_T$, the propagator $\exp(i\vec{p_T}\cdot\vec{b})$ comes into play and the cross-section scales as $\sigma\propto 1-\cos(\vec{p_T}\cdot\vec{b}/\hbar)$, so for $p_T<\hbar/\langle R_A\rangle$, production is suppressed.   
For $\rho$ photoproduction accompanied by mutual Coulomb excitation at RHIC, $\langle\hbar/\langle R_A\rangle \approx 20$ fm, and significant suppression has been observed for $p_T<25$ MeV/c  \cite{Abelev:2008ew}.  

The alignment also causes azimuthal angular correlations between different particles produced in the same ion-ion interaction.  These correlations are visible in the azimuthal angular distributions of the decay products.  The decay products of a linearly polarized spin-1 particle (like a vector meson or giant dipole resonances) form a $\cos(\theta)$ azimuthal distribution with respect to the photon polarization.  Classically, the two linear polarizations follow $\vec{b}$, and the azimuthal angular distribution between the $p_T$ of the particles produced in the two decays is \cite{Baur:2003ar}
\begin{equation}
P(\Delta\theta) = 1 + \frac{\cos(2\Delta\theta)}{2}.
\end{equation}
This neglects the $p_T$ of the spin-1 vector mesons or excited nuclei, a very small correction. These correlations should already be observable at RHIC, in cases like mutual Coulomb excitation to a giant dipole resonance, where both nuclei are excited and decay by single neutron emission; the direction of the $p_T$ of the neutrons can be determined using segmented zero degree calorimeters.  

When two identical vector mesons ({\it e. g.}  $\rho^0\rho^0$) are produced,  there will also be bosonic enhancements.  Since each ion can be produced at either nuclear target, there are four diagrams to produce two $\rho^0$ mesons in a single ion-ion encounter \cite{Klein}. In the simplest case, production at large rapidities, the dominant diagram for double-meson production has both mesons coming from the same nucleus.  Then, the double-$\rho$ production amplitude $A_{\rho\rho} (y)$ is 
\begin{equation}
A_{\rho\rho} \propto A_{\rho}^2(y)\exp(i|\vec{p_{T1}}-\vec{p_{T2}}|R_A/\hbar)
\end{equation}
The probability to produce two identical vector mesons with a small $\vec{p_{T1}}-\vec{p_{T2}}$ is enhanced.  The cross-section for $\rho^0\rho^0$ pair production \cite{Klein:1999qj}  is large enough  that this enhancement should be observable with existing RHIC data.  A more detailed calculation of $\rho^0\rho^0$ photoproduction is given in Ref. \cite{Klusek-Gawenda:2013dka}.

One notable goal for future RHIC UPC studies is to measure the GPD-E helicity-flip parton distribution, which is related to the angular momentum carried by the quarks in a nucleus \cite{Burkardt:2002ks}. The experimental channels is $J/\psi$ photoproduction on polarized proton targets.  This is most feasible using $pA$ interactions where the ion emits the photon  \cite{Elke}.  This is one of the few ways to measure the GPD-E parton distribution.  It may also be possible to use other  final states to study other photoproduction spin effects. 

\section{$pp$ UPCs}

\begin{figure} [tb]
\center{\includegraphics[width=0.7\textwidth]{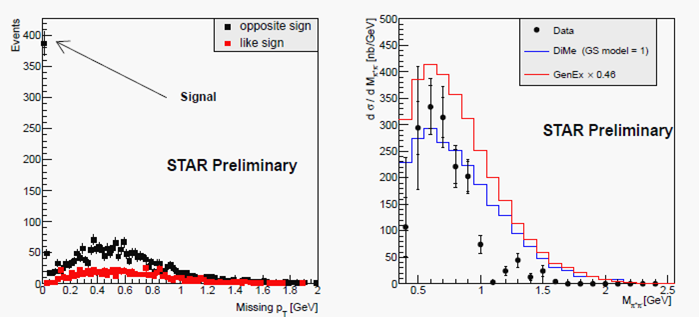}}
\caption{\label{fig:Romanpots} (left) $p_T$ distribution for events with two tracks in the STAR central detector, plus two scattering protons observed in the Roman pots.  The large peak for un-like sign pairs at $p_T\approx 0$ is the exclusive production signal.  The right panel shows the $\pi^+\pi^-$ invariant mass distribution, compared to two Monte Carlo simulations based on double-Pomeron interactions.  From Ref. \cite{Przcien}.
} 
\end{figure}

STAR has recently installed two sets of Roman pots to detect protons that are slightly scattered from the beam.  The initial purpose was to study $pp$ elastic scattering, but they are also used to study $pp$ central diffraction events, whereby the $pp$ collision produces a central state in the detector, with rapidity gaps between that central state and the protons on both sides  \cite{Przcien}.    In previous running, the Roman pot locations were optimized to study elastic scattering, but a small sample of exclusive events was observed; they were selected by requiring that the $p_T$ of the event be balanced within 20 MeV/c, including the proton $p_T$ measured in the Roman pots; this led to a clean selection of exclusive events, with almost no background (as determined from like-sign events).  
 
Figure \ref{fig:Romanpots} shows the $p_T$ distribution of the events; the large peak for $p_T<10$ MeV/c is from fully reconstructed interactions.  The right panel shows the invariant mass distribution, compared with two simulations based on double-Pomeron interactions. There is a hint of $f_2(1270)$ production, as expected in double-Pomeron interactions.  No $\rho$ peak is visible, showing that photon-Pomeron interactions do not play a major role.  However, other photoproduction reactions may still be of interest; $J/\psi$ photoproduction has been observed in $pp$ collisons at the LHC, and in and $p\overline p$ collisions at the Fermilab Tevatron.

\section{Toward an electron-ion collider}

The major physics goal of an $eA$ program at an electron ion collider is to study the internal structure of heavy nuclei, including by studying how the quarks and gluons in nucleons are altered when the nucleons are placed into a heavy nucleus \cite{Accardi:2012qut}.  In $eA$ collisions, the electron emits a virtual photon, which flutuates to a $q\overline q$ pair which then interacts with the nucleus; vector meson photoproduction is one of the main reactions being studied  \cite{Toll:2012mb}.   As with UPCs, $eA$ interactions are sensitive to the nuclear gluon distribution. The main advantage of $eA$ collisions is that one can detect the scattered electron and determine the $Q^2$ of the photon, independent of the hadronic final state.  It is possible to study the gluon distributions at a variety of $Q^2$, in contrast to UPCs, where the $Q^2$ is fixed by the mass of the final state quarks.

\section{Conclusions}

Ultra-peripheral collisions have been used to explore a number of key physics topics at RHIC.  $e^+e^-$ production has been used to study higher order QED and to measure bound-free pair production.  Vector meson photoproduction has been used to study nuclear structure and, particularly shadowing.   RHIC and LHC data on $\rho$ photoproduction show evidence for significant nuclear suppression.  Future data should probe nuclear shadowing in $\rho$ and $J/\psi$ in consideably greater detail.  In the longer term, UPC physics will pave the way for $eA$ analyses at an electron-ion collider.  

\section*{Acknowledgements}

This material is based upon work supported in part by the U.S. Department of Energy, Office of Science, Office of Nuclear Physics, under contract number DE-AC-76SF00098.

\end{document}